%% file: GajaMor.tex
\documentclass{jac}
\usepackage{amssymb,epsfig}

\def\Nset{\mathbb{N}}
\def\Zset{\mathbb{Z}}

\def\Qset{\mathbb{Q}}
\def\ie{{\em i.e.}}

\begin{document}

%----------------opening--------------
\title{Time-symmetric Cellular Automata}
\author[UTFSM]{A. Moreira}{Andr\'es Moreira}
\address[UTFSM]{
Departamento de Inform\'atica and Centro Tecnol\'ogico de Valpara\'iso (CCTVal)
\\ Universidad T\'ecnica Federico Santa Mar\'ia
\\ Casilla 110-V, Valpara\'iso, Chile}
\email{amoreira@inf.utfsm.cl}

\author[UDEC]{A. Gajardo}{Anah\'i Gajardo}
\address[UDEC]{
Departamento de Ingenier\'ia Matem\'atica and Centro de Investigaci\'on en Ingenier\'ia Matem\'atica (CI2MA)
\\ Universidad de Concepci\'on
 \\ Casilla 160-C, Concepci\'on, Chile}

\thanks{This work has been supported by CONICYT FONDECYT \#1090568 (A. Gajardo) and \#1080592  (A. Moreira), as well as UTFSM grant 241016 (A. Moreira).}


\begin{abstract}
  \noindent Together with the concept of reversibility, another relevant physical notion is time-symmetry, which expresses that there is no way of distinguishing between backward and forward time directions. This notion, found in physical theories, has been neglected in the area of discrete dynamical systems. Here we formalize it in the context of cellular automata and establish some basic facts and relations. We also state some open problems that may encourage further research on the topic.
\end{abstract}

\maketitle

\section{Introduction}

\input{introduccion}

\section{Some motivating examples}
\label{sec:ejemplos}

\input{ejemplos}

\section{Basic results}

\input{resultadosB}

\section{Involutions}

\input{involuciones}

\section{Diversity in the class of time-symmetric CA}

\input{ts}

\section{Conclusions}

\input{conclusiones}

\bibliography{Xbib}
\bibliographystyle{plain}

\end{document}

%% file: introduccion.tex
An important property that may be present or not in physical or abstract dynamical systems is reversibility; consequently, it has also been an active topic of research in the context of cellular automata\cite{kariRev}. At least two particular reasons for this interest are often mentioned: on one hand, if CA are seen as models for massive distributed computation, then Landauer's principle suggests that we should focus on reversible cases. On the other hand, reversibility is often observed in real systems; it is therefore desirable in models of them\cite{Toffoli1990229}. Furthermore, a number of interesting results (like the dimension-sensitive difficulty of deciding reversibility\cite{kari2d}) have kept reversible CA in sight over the years.

However, there is one aspect of reversibility, as seen in real systems, which has been mostly neglected when considering cellular automata (in fact, for discrete dynamics in general): the dynamical laws governing physical reality seem to be not only reversible, but {\em time-symmetric}. For Newtonian mechanics, relativity or quantum mechanics, we can go back in time by applying the same dynamics, provided that we change the sense of time's arrow, through a specific transformation of phase-space. In the simplest example, Newtonian mechanics, the transformation leaves masses and positions unchanged, but reverses the sign of momenta.

In the most general sense, we say that a dynamical system $(X,T)$ is time-symmetric if there exists a reversible $R:X\to X$ such that $R\circ T\circ R^{-1} = T^{-1}$~\cite{Lamb}(notice that this applies to systems with discrete or continuous time). However, time-symmetries observed in physical systems follow usually a more restricted definition, in which $R^{-1}=R$, and therefore $R$ is an {\em involution} on $X$. This is a natural restriction, which follows whenever there is no way to distinguish where the arrow of time is heading. Apparent irreversibility (Loschmidt's paradox) comes only from macroscopic (\ie, coarse-grained) differences in entropy.

Here we will discuss some basic facts about time-symmetric cellular automata, defined as those CA $F$ for which there exists an involution H (which is a CA itself) such that
\begin{equation}
 F^{-1} \;=\; H \circ F \circ H   \label{eq:condicion}
\end{equation}
Requiring $H$ to be a CA is somewhat arbitrary, since for other systems the time-reversing transformation is not necessarily of the same nature as the dynamics (in fact, the physical theories discussed above are continuous in time). The reason for this restriction is that we expect reversibility (including the particular case of time-symmetry) to be a {\em local} property. Even if we do not address the case when $H$ is not a CA, it may be an interesting direction for future studies.

%%%%%%%%% 
CA are usually defined over a full shift $S^\Zset$, but they can also be studied over (stable) subshifts. We remark that in this case, for time-symmetry to apply, the subshift must be stable for both $F$ and $H$. This may cause some problems, since subsystems of a time-symmetric CA cannot be assumed to be time-symmetric too, even if they are stable for $F$.
%because we cannot assert that subsystems of time reversible CA are also time-reversible, because closeness for $F$ does not implies closeness for $H$, even in the time-reversible case.
%%%%%%%%%%%%%

%% file: ejemplos.tex
Not only our models of physical reality turn out to exhibit time-symmetry; it is also found in some well known reversible discrete dynamical system. We show in this section how it applies to two 2D system, Margolus' billiard and Langton's ant. As a technical note, notice that in both cases the system is not originally described as a cellular automaton in the strict sense; therefore, we describe for each of them a CA that contains them as particular case for a subshift of valid configurations. This should -in principle- be followed by an extension of the rule to the full shift, and such that the system remains time-symmetric; however, doing that is not really required, since we want to show the time-symmetry of the original system; we hence restrict ourselves to the valid subshift.

\begin{figure}[htb!] 
 \begin{center}
  \begin{tabular}{cp{1cm}cp{1cm}c}
   \resizebox{4cm}{!}{\includegraphics{BilliardPart.mps}} & &
   \resizebox{2.5cm}{!}{\includegraphics{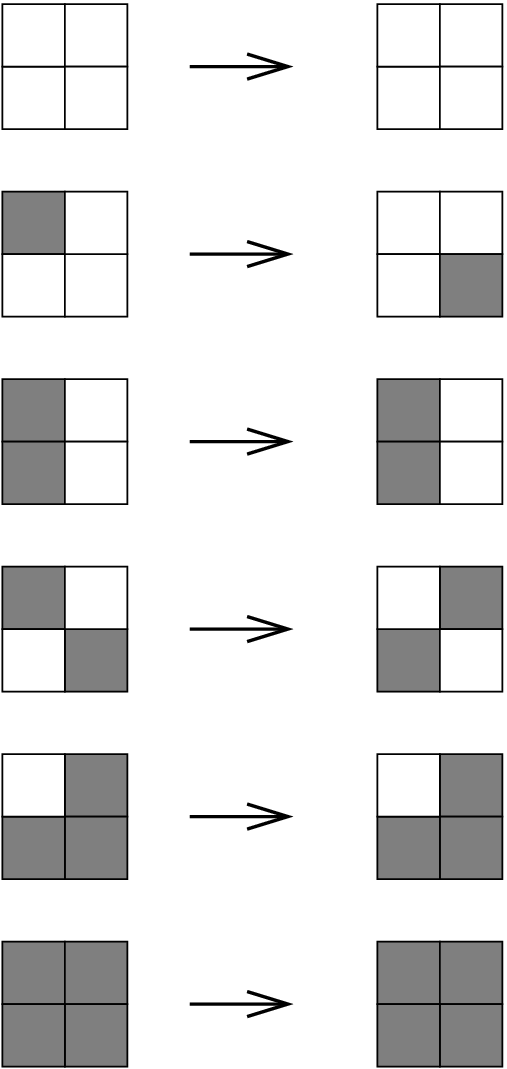}} & &
   \resizebox{4cm}{!}{\includegraphics{BilliardArrow.mps}} \\
   (a) && (b)&& (c)\\
  \end{tabular}
  \caption{(a) The two partitions of the Margolus model are shown, one with solid lines and the other with dashed lines. (b) The Billiard Ball Model is defined through a permutation over $2\times 2$ blocks of cells. (c) The current partition is obtained by grouping the four cells that point to the same point; reversing the arrows gives the alternate partition.}
  \label{fig:margolus}
 \end{center}
\end{figure}

{\bf Margolus' billiard. }
A well known example of time-symmetric CA is the Billiard ball model of Margolus~\cite{Marg}.
It is not a proper CA, but rather a so-called {\em partitioned CA}, where the space $\Zset^2$ is partitioned in $2\times 2$ blocks of cells in two different ways (see Figure \ref{fig:margolus}(a)).
A transformation is applied to each block of each partition alternately.
It is easy to see that such an automaton is reversible if and only if its local transformation is one-to-one.
The rule used by Margolus is shown in Figure \ref{fig:margolus}(b).
It tries to emulate balls that move in straight lines, colliding elastically with each other or with static obstacles.
The importance of this model comes from its Turing-universality, proved in \cite{Marg} by computing reversible Fredkin gates \cite{Fred82}.

We can express Margolus' system in terms of a CA with alphabet $\{white,black\}\times\{\nearrow,\searrow, \swarrow, \nwarrow\}$ and Moore neighbourhood. Here the first layer (the white/black component) represents the states of the original Margolus model, and the other represents the current partition, along with the relative position of the cell within its current block.
This layer must be initialized in an appropriate way in order to work correctly (see Figure~\ref{fig:margolus})(c). 

Notice that reversing the arrows makes the partition flip to the alternative one. At each time step, each cell computes its next white/black state by applying Margolus' rule to the quadrant indicated by its arrow, and then reverses its arrow. Each of this actions -the first on the first layer, the second on the second- is an involution. Furthermore, if at one time step we omit any of them, further iterations will make the automaton evolve back in time.

%DECIDIR QUÉ PASA CUANDO LA ci NO ES LA CORRECTA DE MANERA QUE EL SISTEMA SIGA SIENDO T-S

\begin{figure}[htb!] 
 \begin{center}
 \begin{tabular}{cp{1cm}c}
  \resizebox{3cm}{!}{\includegraphics{AntRule.mps}}  &&
  \resizebox{6cm}{!}{\includegraphics{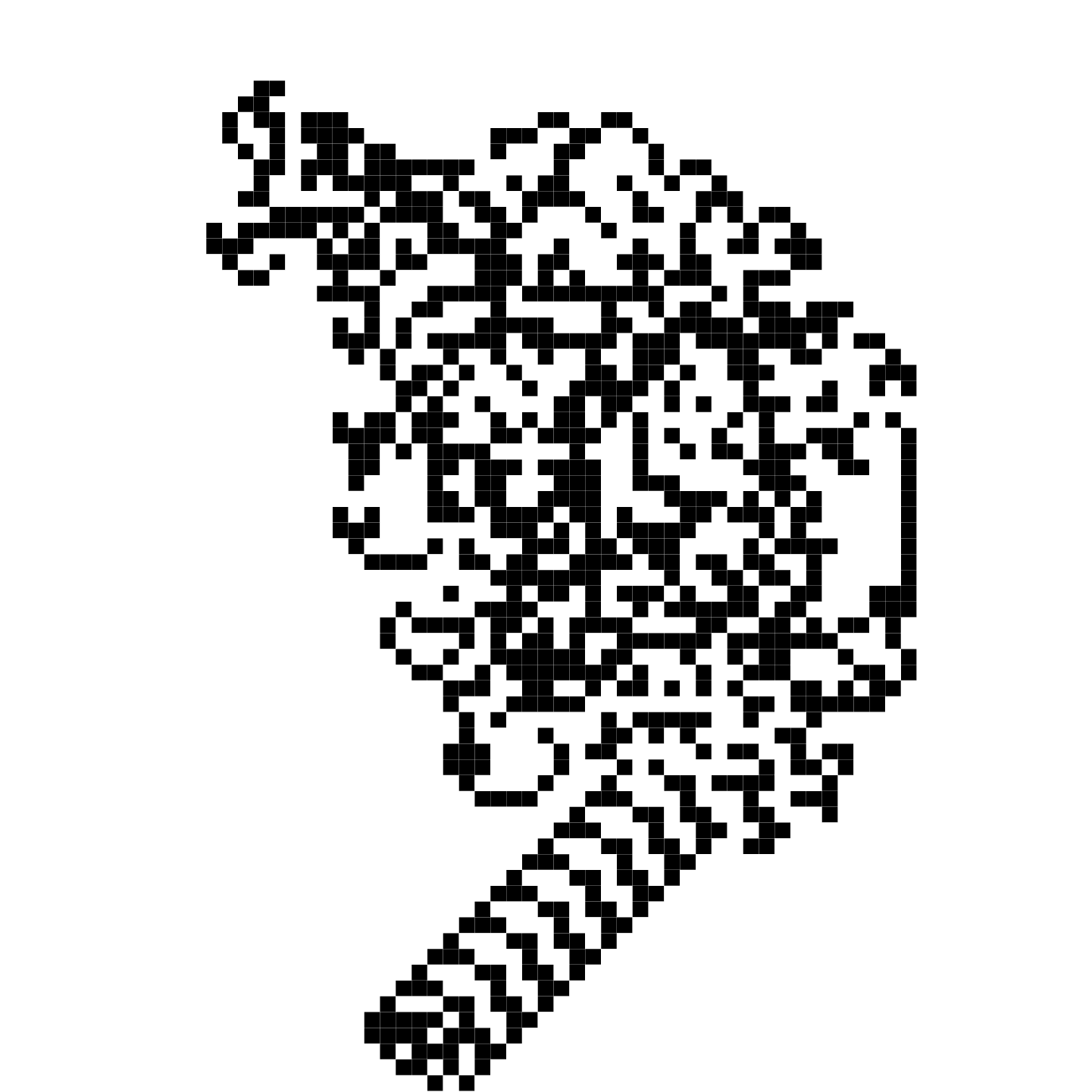}} \\
  (a) && (b)\\
 \end{tabular}
 \caption{(a) Langton's ant rule. (b) Space configuration at iteration 10,837 after starting with every cell in \emph{white} color.}
  \label{fig:ant}
 \end{center}
\end{figure}

{\bf Langton's ant. }
Langton's ant was introduced in~\cite{Lang} together with several models emulating different life properties.
It was also defined in physics as a model for particles presenting self correlated trajectories \cite{Cohe92}.
The model can be seen as a Turing machine working on a 2-dimensional tape.
Its internal state is an arrow that represents its last movement direction.
At each step, the ant turns to the left or to the right depending on the cell color (\emph{white} or \emph{black}), it flips this color and moves one cell forward (see Figure \ref{fig:ant}(a)).
Besides being Turing-universal \cite{nosotros}, its celebrity is due mostly to its particular behavior over finite initial configurations.
Simulations show that it always falls eventually into a repetitive movement -of period 104- that makes it propagate unboundedly (see Figure \ref{fig:ant}(b)); this assertion has not been proved, and appears to be very difficult despite the simplicity of the transition rule.

Langton's ant can also be described in terms of a CA with Moore neighborhood and state cell $\{head,tail,empty\}\times\{white,black\}$.
We represent the arrow through two adjacent cells, one in state \emph{head} and the other in state \emph{tail}.
The cell in state \emph{tail} always becomes \emph{empty}, while the cell in state \emph{head} always becomes \emph{tail} and flip its color.
Cells adjacent to a \emph{head} can decide to become \emph{head} themselves by looking at the tail position and the color of the head cell.
The system simulates Langton's ant only if it starts with only one ant.

Here again, we can define the involution consisting in exchanging \emph{tails} and \emph{heads}.
This immediately makes the ant come back to the cell it just had left, which it finds in the color opposite to the one it had found before, causing the ant to turn in the opposite direction, which in turn makes it again go to a previously visited cell, and so on: the ant will forever retrace (and undo) its past trajectory.

%\begin{figure}[htb!] 
% \begin{center}
%  \begin{tabular}{cp{1cm}c}
%   \resizebox{3cm}{!}{\includegraphics{AntRule}} &&
%   \resizebox{3cm}{!}{\includegraphics{AntRevertion}} \\
%   (a) && (b)\\
%  \end{tabular}
%  \caption{(a) Langton's Ant Rule. (b) Reverting the arrow reverts the time direction.}
%  \label{fig:AntRule}
% \end{center}
%\end{figure}

%% file: resultadosB.tex
\begin{prop} 
Let $F$ be a CA. Then the following are equivalent:
\begin{enumerate}
  \item
   $F$ is time-symmetric.
  \item
   There exists an involution $H$ such that $(F\circ H)$ is an involution.
  \item
   $F$ is the composition of two involutions.
 \end{enumerate}
\end{prop}

\proof

 (1)$\implies$(2) \\
  Let $F$ and $H$ be the CA satisfying \ref{eq:condicion}. Then
  \[
  (F\circ H)^2 \;=\; F\circ H \circ F\circ H \;=\; F\circ F^{-1} \;=\; id
 \]

(2)$\implies$(3) \\
  Take $H$ from (2) and let $G=F\circ H$ which is an involution. We have
  \[
   F \;=\; F\circ id \;=\; F\circ (H\circ H)  \;=\; (F\circ H) \circ H \;=\; G\circ H
  \]

(3)$\implies$(1) \\
  Let $G$ and $H$ be involutions such that $F=G \circ H$. Then
  \[
   F^{-1} \;=\; (G \circ H)^{-1} \;=\; H^{-1}\circ G^{-1} \;=\; H \circ G 
	  \;=\; H \circ G \circ H \circ H \;=\; H \circ F \circ H
  \]
 \qed

\begin{rems}
The following additional facts are noteworthy:
\begin{enumerate}
 \item 
  If $F$ is time-symmetric, then so is its inverse $F^{-1}$. Moreover, if $F=G\circ H$ is a decomposition into involutions, then $F^{-1}=H\circ G$ is a decomposition for the inverse. If $H$ was the involution verifying \ref{eq:condicion}, then $G$ plays that role for $F^{-1}$.
 \item
  For any $i\in \Zset$, $F^i$ is also time-symmetric.
 \item 
  The identity is a (trivial) involution; from there and the third condition we have that any involution is trivially time-symmetric.
 \item
  Not every reversible CA is time-symmetric. For example, $\sigma$ (the shift): if for some $H$, $(\sigma\circ H)\circ(\sigma\circ H)=id$, since any CA commutes with the shift, we would have $\sigma^2=id$, which is a contradiction.
\end{enumerate}
\end{rems}

%... this can be discussed, by taking a reflection as involution we reverse the shift...etc..

The following diagram commutes:

\[\renewcommand{\arraystretch}{1.5}
\begin{array}{ccc}
X  & \begin{array}{c} \xrightarrow{\quad h\quad } \\ \xleftarrow[\quad h\quad]{} \end{array} & X \\
F \Big\downarrow\Big\uparrow {F^{-1}} &  & F \Big\downarrow\Big\uparrow {F^{-1}} \\
X  & \begin{array}{c} \xrightarrow{\quad h\quad } \\ \xleftarrow[\quad h\quad]{} \end{array} & X \\
\end{array}\]

\begin{figure}[htb!] 
 \begin{center}
   \resizebox{6cm}{!}{\includegraphics{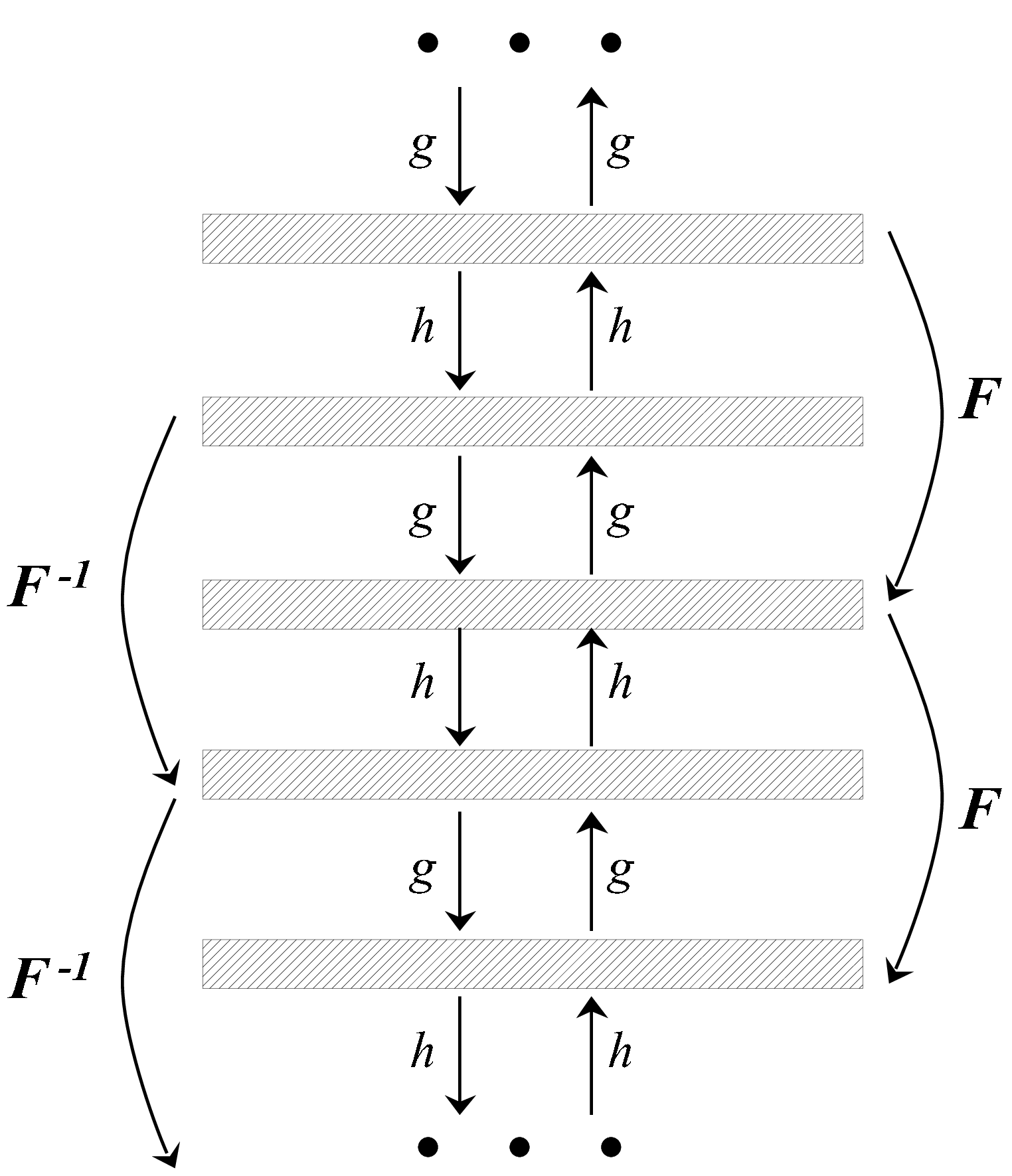}}
  \caption{The decomposition of time-symmetric CA into alternating involutions creates a situation where both $F$ and its inverse can be read from the space-time diagram as time moves forward (or backward).}
  \label{fig:diag}
 \end{center}
\end{figure}

Moreover, if we use $F=G\circ H$ to decompose the dynamics into the alternate applications of the involutions, so that successive configurations are computed as $c'_t=H(c_t)$, $c_{t+1}=G(c'_t)$, we get a dynamics $c_0$, $c'_0$, $c_1$, $c'_1$, \ldots, where both $F$ and $F^{-1}$ are being iterated: $c_{t+1}=F(c_t)$ and $c'_{t+1}=F^{-1}(c'_t)$. This curious situation is represented in Figure \ref{fig:diag}.

%% file: involuciones.tex
Involutions are quite infrequent in the space of CA. For example, the only elementary CA of period two are the identity and its negation.

It is easy to decide whether a given CA is an involution or not: we just have to compute its square and compare it to the identity.
Nevertheless, if we want to enumerate the set of involutions, this procedure is very slow. A constructive characterization or a practical set of strong necessary conditions is still missing.

Meanwhile, it may be useful to consider restricted families of CA. For instance, if we restrict ourselves to additive CA, we get an alternative characterization in terms of coefficients. If we consider an additive CA of radius $r$ defined on $(\Zset_m)^\Zset$ by the local rule
\[
 h(x_{-r},\ldots,x_0,\ldots,x_r) \;=\; \sum_{i=-r}^r a_i x_i
\]
then by applying the rule twice and grouping the terms it is easily seen that $h$ is an involution if and only if
\[
 a_0 + 2 \sum_{i=1}^r a_i a_{-i} = 1  \quad \textrm{and} \quad  \sum_{i=-r+\max\{j,0\}}^{r+\min\{j,0\}} a_i a_{j-i} = 0 \, , \forall j\neq 0
\]
For instance, for $m=4$ and $r=1$, we get $2 a_{-1} a_0 = 2 a_0 a_1 = a_{-1}^2 = a_1^2 = 0$ and $a_0^2+2a_{-1}a_1=1$. Putting $a_{-1}=a_1=2$ we obtain all the zeroes, and with $a_0=1$ or $a_0=3$ we have the last condition. An example of an additive involution is thus
\begin{equation}
 h( x_{-1},x_0,x_1 ) \;=\; \left( 2(x_{-1}+x_1) + 3 x_0 \right) \mod 4   \label{eq:invo1}
\end{equation}

Another well studied family of CAs are permutative ones. Unfortunately, we do not have a characterization there. A necessary condition is given by the following fact:

\begin{prop}
Given an involution $h$, the following two assertions are equivalent:
\begin{itemize}
\item $h$ is left-permutative.
\item $h$ is one-way to the right\footnote{The neighbourhood is a finite subset $N$ of $\Nset_0$.}. %, the set of positive or integers together with 0
\end{itemize}
\end{prop}
\proof
If $h$ is left-permutative of left radius $l$, $h^2$ is also left-permutative of left radius $2l$, but the left radius of $h^2$ is 0, then $l$ is 0.
Conversely, if $h$ is oneway to the right and it is not permutative, there exists $x_1...x_n$ and $y_1$ such that $h(x_1,...,x_n)=h(y_1,x_2,..,x_n)$.
Taking any extension $z$ of $x_1..x_n$ to $\Zset$, and $z'_i=z_i$ for every $i\not=1$ and $z'_1=y_1$, we have that $h(z)_{[1,\infty[}=h(z')_{[1,\infty[}$.
Thus, $z'_1=h^2(z)_1=h^2(z')_1=z_1$, which is a contradiction.
\qed

Thus the involution in (\ref{eq:invo1}), which is clearly two-way, is not permutative. An example of a permutative involution of radius $r$ is:

\begin{equation}
h(x_{-r},..x_0)=\left\{\begin{array}{ll}
1&\textrm{if } x_{-r}=0 \wedge x_0=2\\
2&\textrm{if } x_{-r}=0 \wedge x_0=1\\
x_0 & \textrm{otherwise}
\end{array}\right.
\label{eq:invo2}
\end{equation}

This kind of construction is the simplest one, with permutations which are transpositions and which do not affect the states that regulate their application. More complicated examples may have associated permutations which are not transpositions.

Example (\ref{eq:invo2}), incidentally, shows how an involution can have an arbitrarily large neighbourhood. This long-distance dependence may be lost when it is composed with another involution, making the determination of time-symmetry non-trivial. For instance, the involution (\ref{eq:invo2}) yields the permutation $(12)$ of radius 0 when it is composed with 

\begin{equation*}
g(x_{-r},..x_0)=\left\{\begin{array}{ll}
1&\textrm{if } x_{-r}\neq 0 \wedge x_0=2\\
2&\textrm{if } x_{-r}\neq 0 \wedge x_0=1\\
x_0 & \textrm{otherwise}
\end{array}\right.
\end{equation*}

%Another interesting property of involutions are their Welch indices.... blabla.

%% file: ts.tex
One simple example of time-symmetric automata is given by the following.

\begin{prop}
Every reversible CA of radius 0 is time-symmetric.
\label{prop:radio0}
\end{prop}

\proof 
Let $f$ be the local rule of a reversible CA of radius 0, and let $S$ be its set of states. Suppose first that
$f:S\rightarrow S$ is a cyclic permutation and, without loss of generality, that $S=\{0,..,n-1\}$ and $f(i)=i+1 \mod n$.
Let us define the involution $h(i)=n-i-1$. Consider $g=h\circ f$; for $i<n-1$, $g(i)=h(f(i))=h(i+1)=n-(i+1)-1=n-i-2$, while otherwise $g(n-1)=h(f(n-1))=n-1$. Thus $g$ is an involution: $g^2(n-1)=g(n-1)=n-1$, and for other $i$, $g^2(i)=g(n-i-2)=n-(n-i-2)-2=i$.
Since $f=h\circ h\circ f$=$h\circ g$ is the composition of two involutions, it is time-symmetric.

If $f$ decomposes into more than one cycle, we define $h$ and $g$ as before over each of them, obtaining again a decomposition into involutions.
\qed

It is important to notice here the preservation of time-symmetry under conjugacy.

\begin{prop}
If $F$ is conjugated to $T$ and $T$ is time-symmetric, then $F$ is also time-symmetric.
\end{prop}
\proof
From time-symmetry, there is an involution $H$ such that $T^{-1}=H\circ T\circ H$. From conjugacy, there is a bijective, continuous, shift-commuting $\phi$ such that $T=\phi \circ F \circ \phi^{-1}$. Then we have (removing the composition symbol, for clarity) that
\[
F^{-1} \;=\; \phi^{-1} T^{-1} \phi  \;=\; \phi^{-1} H T H \phi \;=\; \phi^{-1} H \phi F \phi^{-1} H \phi \;=\; G F G
\]
and $G=\phi^{-1} H \phi$ is cleary an involution, making $F$ time-symmetric.
\qed

%%%%%%%%%%%%%%%%%%%%%%
Periodic CA of radius $r>0$ behave almost like a CA with radius 0, in the sense that information cannot travel ``very far''; this makes them nearly time-symmetric, because they are conjugated to a subshift of a radius 0 CA. To see this, let $F$ be a $p$-periodic CA with states $S$ and define $\varphi:S^\Zset\to (S^p)^\Zset$ as $\varphi(x)_i=(x_i,F(x)_i,..,F^{p-1}(x)_i)$.
This $\varphi$ is continuous and injective, and the induced CA $F'$ in $(S^p)^\Zset$ has radius 0 and period $p$; its local rule is $ f'(a_0,a_1,..,a_{p-1})=(a_1,a_2,..,a_{p-1},a_0) $.
From Proposition \ref{prop:radio0} we see that $F'$ is time-symmetric; moreover, $\varphi(S^\Zset)$ is $F'$ invariant. However, $\varphi(S^\Zset)$ is not invariant for the involution defined in the proof, and therefore we cannot conclude that $F$ is time-symmetric.
%%%%%%%%%%%%%%%%%%%%

\begin{figure}[htb!]
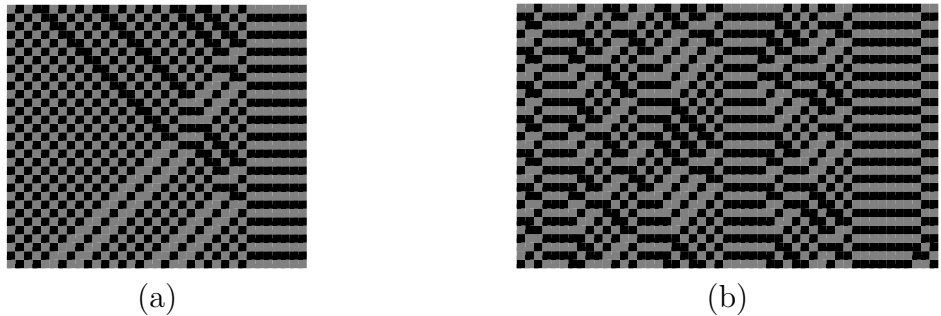
 
 \begin{center}
   \begin{tabular}{cp{2cm}c}
   \resizebox{4cm}{!}{\includegraphics{InfiniteOrder.mps}} & &
   \resizebox{5.6cm}{!}{\includegraphics{Random.mps}} \\
   (a) & & (b) \\
   \end{tabular}
  \caption{Simulations of the time-symmetric CA defined by Hedlund \emph{et al} on periodic boundary conditions.
  (a) Two traveling signals.
  (b) A simulation over a random initial configuration.}\label{fig:Hedlund}
 \end{center}
\end{figure}

When we consider the group of reversible CA with the composition operation, involutions correspond to elements of order two. These objects were already considered by Hedlund {\it et al} in~\cite{Hedl}, where the last theorem shows that the composition of two involutions (i.e., time-symmetric CA) can have infinite order.
The example that proves this theorem  is defined on alphabet $\{0,1\}$ and consists in the composition of $\alpha$, the negation of the identity, and $\beta$, a CA  that negates $x_i$ if and only if $x_{i-1}x_{i+1}x_{i+2}=101$.
Figure~\ref{fig:Hedlund} shows simulations of this CA; the Hedlund's proof exhibited a configuration with infinite orbit consisting in one traveling signal over a periodic background like the one appearing in Figure~\ref{fig:Hedlund}(a). 

This already suggests a variety of dynamical behaviors within the class. 
But the examples given in Section \ref{sec:ejemplos} are even more interesting, as they correspond to Turing-complete systems.
The following results shows that, indeed, the whole range of reversible dynamical behaviors can be observed in time-symmetric CA.

\begin{prop}
Let $F$ be a 1D reversible CA. Then there exists a 1D CA $\tilde{F}$ which is time-symmetric and simulates $F$ in real time.
\end{prop}

\proof
Let $f$ be the local rule of $F$ and denote with $f^{-1}$ the local rule of its inverse $F^{-1}$; let $\ell$ and $r$ be large enough so that $N=\{-\ell,\ldots,r\}$ contains the neighbourhoods of both $f$ and $f^{-1}$; finally, let $S$ be the set of states. We define the CA $\tilde{F}$ with neighbourhood $N$ and states $S^2$, through the local rule
\[
 \tilde{f} \left( (x_{-\ell},y_{-\ell}),\ldots, (x_r,y_r) \right) \;=\;
 ( f(x_{-\ell},\ldots,x_r) , f^{-1}(y_{-\ell},\ldots,y_r) )
\]
$\tilde{F}$ simulates $F$ in real time: to project the space-time diagram of $\tilde{F}$ into that of $F$, we just discard the second component of the ordered pairs. By discarding the first component instead, we note that $\tilde{F}$ simulates $F^{-1}$ as well.

Let $H$ be the involution given by the radius 0 local rule $h(x,y)=(y,x)$. Abusing notation, denote configurations $c\in (S^2)^\Zset$ as pairs $(x,y)\in (S^\Zset)^2$. Then we have
\[
 \tilde{F}\circ H (x,y) \;=\; \tilde{F} (y,x) \;=\; (F(y),F^{-1}(x))
\]
and
\[
 (\tilde{F}\circ H)^2 (x,y) \;=\; \tilde{F}\circ H (F(y),F^{-1}(x)) \;=\; \tilde{F} (F^{-1}(x),F(y)) \;=\; (x,y)
\]
and thus $\tilde{F}$ is time-symmetric.
\qed

Cellular automata are said to be intrinsically universal if they are able to simulate any other CA. The details vary according to the accepted notion of \emph{simulation}, from which there is a variety. Delorme \emph{et al}~\cite{Mazo10II} have recently reviewed and completed the study of three of these, \emph{surjective, injective} and \emph{mixed} simulation, and shown that for every pair of CA $F$ and $G$, $F\times G$ simulates both $F$ and $G$ in all three senses.

\begin{cor}
There exist time-symmetric CA which are intrinsically universal within the class of reversible CA.
\end{cor}

\proof
This follows from the previous results and comment, and from the existence of reversible intrinsically universal CA (see for example \cite{ollinger_jac_2008}).
\qed

Notice that reversible CA cannot simulate arbitrary CA: intrinsic universality is therefore limited to the reversible class, and time-symmetric CA are as general as reversible CA can get. Turing-universality is not limited by reversibility (information can be ``swept away'' to preserve it and maintain reversibility) and hence is implied by reversible intrinsic universality.

%%%%%%%%%%%%%%%%%%%%%%%%%%
Not every reversible CA is time-symmetric; a simple example is the shift $\sigma$, which commutes with every CA and therefore cannot satisfy equation (\ref{eq:condicion}) for any involution $H$.
But in general, it is not easy to prove non-time-symmetry.
An interesting theory which may provide better tools for doing this, and possibly for characterizing time-symmetry, is the one developed by Kari in \cite{Kari96}. We will not reproduce here his construction, but one important fact is the following: he introduces a morphism $h_-$ from the set of reversible CA with the composition $(Aut(A),\circ)$ into the set of rational numbers with the multiplication $(\Qset, \cdot)$: $h_-(f\circ g)=h_-(f)h_-(g)$ for all reversible $f,g$. Clearly, involutions and every periodic CA are in the kernel of $h_-$. Moreover, since time-symmetric CA are compositions of involutions, they are in this kernel as well. We do not presently know whether they are identical to the whole kernel or not. 

Kari proves that reversible CA which are not in this kernel are compositions of some element of the kernel with a \emph{partial shift} which is easily computed from the value of $h_-$; in turn, every element of the kernel can be written as a composition of two block permutations (akin to Margolus rule), and thus he expresses reversible CA in an explicitly reversible way. Our motivation here is different, but the approach is promising and the connection should be explored. As a first conclusion, we obtain that every CA in the kernel of $h_-$ is a composition of two time-symmetric CAs, and hence is also the composition of four involutions.
%%%%%%%%%%%%%%%%%%%%%%%%%%

%% file: conclusiones.tex
We believe that this note just scratches the surface of the topic of time-symmetry in cellular automata; their rich internal structure and the connection to physical models suggests that much more can be done with them.

On the other hand, as shown by the examples in Section \ref{sec:ejemplos}, time-symmetric CA are actually quite familiar to CA researchers, and have appeared in different contexts. Some cases are very explicit, like the automata constructed in the proof of undecidability of periodicity \cite{KariInmortal}, which actually include an ``arrow of time'' toggle. Moreover, there are ways of constructing CA rules that make the construction of time-symmetric CA straightforward. 
For instance, Margolus' billiard is an example of a block automata, {\it i.e.}, a system which is a composition of two functions applied to independent blocks of the configuration. By incorporating the current function and block to be applied into the configuration, a block automaton can always be expressed as a CA. Defining an involution that toggles the current block is a good idea to prove time-symmetry, but it only works if both block functions are also involutions. What must be stressed is that this is only a {\em sufficient} condition; the system may be time-symmetric by means of an entirely different involution.

Likewise, partitioned CA (in the sense of Morita~\cite{MoriParti}) can easily give birth to time-symmetric CA. In that case, cells are partitioned into sub-cells, one for each neighbours; iteration proceeds by the alternation between an exchange step, where cells exchange the contents of the sub-cells associated to each other, and a step which applies a block transformation on the cell. This scheme was succesfully used to construct reversible CA (all we need is a reversible block transformation), and can produce time-symmetric CA as well if the block transformation is chosen as an involution: the exchange step already is one. Again, what we want to stress is that this is a sufficient condition: we could have a partitioned CA which is time-symmetric while having a non-involutive block transformation, if the decomposition happens to be another one.

There are several interesting questions that should probably be addressed next, and have appeared along this text:
\begin{itemize}
 \item
  Is there a constructive characterization of CA involutions that can make their enumeration practical? Right now the only way we have to find the involutions is to test all CA exhaustively; some trivial necessary conditions can be used to reduce the search, but they are not enough to make it efficient.
 \item
  Is time-symmetry a decidable property? Since the definition calls for the existence of an involution that verifies a condition, a bound on the necessary neighbourhood for the involution would be enough to ensure decidability.
 \item
  Do time-symmetric CA correspond to the kernel of Kari's $h_-$ morphism?
\end{itemize}

We conjecture a positive answer for these three questions, at least in dimension 1; Kari's result on undecidability of reversibility in dimension 2 \cite{kari2d} suggests that answers here may be dimension-sensitive too. The answers to the questions may be related to each other. For instance, a better understanding of the structure of involutions may be useful for bounding the required neighbourhood and thus deciding time-symmetry. On the other hand, since $h_-$ is easily computed, a positive answer to the third question would imply a positive answer to the second as well.%%%%%%%%%%%%%%%%%%%%%%%%%%

Notice that if the answer to the third question is positive, then time-symmetric CA would be closed under composition. This is by no means obvious, and in fact it is a further interesting open question.
%%%%%%%%%%%%%%%%%%%%

A further direction for future work may be the study of time-symmetry in other discrete dynamical systems. In each case an important issue is to precise what kind of involution is to be applied. Generally speaking, what we need is an involutive and hopefully local transformation of the system's configuration. That transformation may not be, in general, an object of the same kind as the dynamics itself: that was the case for CA because of the special nature of CA, which transform the whole configuration in discrete time too, and will be the case for automata networks in general. In other cases, like for instance Turing machines, it is not only difficult (the composition of two Turing machines moves the head two steps, and is no longer a Turing machine unless we extend the definitions) but also not expected; rather, for Turing machines, the involution would likely be a transformation on the tape (a CA involution?) along with a change in the current state of the machine. Finally, the locality of the time-reversing involution is not completely granted either: even in CA, it would be interesting to see what happens if that requirement is removed.

\section*{Acknowledgment}
Andr\'es Moreira thanks Gaetan Richard for useful remarks on the characterization of time-symmetry, and Nicolas Ollinger and the Laboratoire d'Informatique Fondamentale de Marseille for the visit during which that discussion took place.